\pdfoutput=1

\documentclass[preprint,onecolumn]{revtex4-2}

\usepackage{graphicx}
\usepackage{siunitx}
\usepackage[hidelinks]{hyperref}
\usepackage{gensymb}
\usepackage{float}
\usepackage{lineno}
\usepackage{xcolor}
\usepackage{amsmath, amsthm, amssymb}

\usepackage[british]{babel}

\usepackage{color}
\usepackage{tabularray}

\begin{document}

\title{Neural network analysis of neutron and X-ray reflectivity data: Incorporating prior knowledge for tackling the phase problem}

\author{Valentin Munteanu\textsuperscript{a}}
\author{Vladimir Starostin\textsuperscript{a}}
\author{Alessandro Greco\textsuperscript{a}}
\author{Linus Pithan\textsuperscript{a}}
\author{Alexander Gerlach\textsuperscript{a}}
\author{Alexander Hinderhofer\textsuperscript{a}}
\author{Stefan Kowarik\textsuperscript{b}}
\author{Frank Schreiber\textsuperscript{a}}

\affiliation{
[a] Institute of Applied Physics, University of Tübingen, Auf der Morgenstelle 10, 72076 Tübingen, Germany
}
\affiliation{
[b] Department of Physical Chemistry, University of Graz, Heinrichstraße 28, 8010 Graz, Austria
}

\begin{abstract}

    Due to the lack of phase information, determining the physical parameters of multilayer thin films from measured neutron and X-ray reflectivity curves is, on a fundamental level, an underdetermined inverse problem. This so-called phase problem poses limitations on standard neural networks, constraining the range and number of considered parameters in previous machine learning solutions. To overcome this, we present an approach that utilizes prior knowledge to regularize the training process over larger parameter spaces. We demonstrate the effectiveness of our method in various scenarios, including multilayer structures with box model parameterization and a physics-inspired special parameterization of the scattering length density profile for a multilayer structure. By leveraging the input of prior knowledge, we can improve the training dynamics and address the underdetermined ("ill-posed") nature of the problem. In contrast to previous methods, our approach scales favorably when increasing the complexity of the inverse problem, working properly even for a 5-layer multilayer model and an N-layer periodic multilayer model with up to 17 open parameters.
    
\end{abstract}

\maketitle
\renewcommand\thefigure{\arabic{figure}}
\renewcommand\thetable{\arabic{table}}

\section{Introduction}

X-ray and neutron reflectometry (XRR and NR) are well-established and indispensable experimental techniques commonly used to investigate the scattering length density (SLD) profile along the direction perpendicular to the surface of a sample such as thin films and multilayers \cite{Tolan1999X, Holy1999High, Sinhaa, Daillant2009X, Zhou1995Theoretical}. The most common way of modeling the SLD profile of a measured sample is via the box model parameterization, where the physical parameters of interest are the thickness, the roughness, and the constant SLD of each layer in a multilayer structure. Still, more complex parameterizations of the SLD profile can be employed based on preexisting physical knowledge or intuition about the investigated structure. XRR and NR have been extensively used in both \textit{in situ} and \textit{ex situ} studies of a large variety of systems, such as liquid and solid thin films \cite{Kowarik2006Real, Woll2011Quantitative, Braslau1988Capillary, Metzger1994Novel, Michely2004Islands, Lehmkuehler2008Carbon, Seeck2002Observation, FragnetoCusani2001Neutron, Festersen2018, Schlomka1996, Treece2019Optimization}, layers of polymers \cite{Ankner1993Neutron, Mukherjee2002Reversible}, lipids \cite{Neville2006Lipid, Skoda2017Simultaneous, Salditt2016, Sironi2016}, self-assembled monolayers \cite{Wasserman1989structure, Skoda2022Switchable} and organic solar cells \cite{Tidswell1990X, Fenter1997Thermally, Lorch2015Growth}. In addition, polarized NR\cite{Majkrzak1991Polarized} can be used to study the magnetic properties of thin films, but it is important to realize that these successful uses of reflectometry usually involve some form of complementary information in the data analysis, such as typical densities or reasonable intervals for the film thickness, provided by the experimentalist.

Reflectometry can be counted among the techniques affected by the phase problem. In the absence of complementary information, the lack of phase information introduces a degree of ambiguity when trying to reconstruct the SLD profile of the investigated sample from the measured reflectivity curve, i.e. the inverse problem is, on a fundamental level, underdetermined ("ill-posed"). This intrinsic ambiguity of the scattering method is further magnified by additional ambiguity due to the limited experimental accuracy, such as instrument noise, measurement artifacts, and the finite number of measured points over the domain of the momentum transfer vector $q_z$.

In recent years, machine learning has emerged as an alternative to classical methods of analyzing surface scattering data \cite{Hinderhofer2023}, being attractive due to its very fast prediction times and the ability to be incorporated into the operating pipelines of large-scale measurement facilities. In particular, fast machine learning-based solutions are ideal for enabling an experimental feedback loop during reflectometry measurements performed in real time \cite{Pithan2023, Ritley2001}. While many machine learning approaches dedicated to reflectivity exist \cite{Greco2019Fast, Greco2021Neural, Greco2022Neural, Mironov2021Towards, Doucet2021Machine, Aoki2021Deep, Kim2021Probabilistic, Andrejevic2022Elucidating}, most of them do not address the inherent ambiguity of the reflectivity data directly, instead training neural networks over specific parameter domains where the ambiguity is not prominent enough to prevent convergence. Such networks lack the flexibility to be used in more general scenarios, typically requiring to be retrained on new parameter domains for each use case. Specifically, Ref. \cite{Greco2022Neural}, while successful within its target, only addresses the case of a layer grown on top of a silicon/silicon oxide substrate.

In our method proposed here, we enhance the regularization of the inverse problem by including prior boundaries for the parameters as supplementary inputs to the neural network. This allows the network to be trained over wide parameter domains while the regression is being conducted over small enough subdomains defined by the prior bounds to mitigate ambiguity. During inference, the output of the network is defined not only by the measured reflectivity curve but also by the prior experimental knowledge for the considered physical scenario. Different choices of prior bounds can lead to the recovery of distinct solution branches from the larger parameter domain. 

Another limitation of existing approaches is that they require a specific discretization of the reflectivity curves, as imposed by common network architectures. While experimental curves can be interpolated to the required discretization, interpolation is prone to introduce unphysical artifacts, for example around deep minima of Kiessig fringes in NR / XRR curves. To address this limitation, we introduce the use of a neural operator for processing reflectivity curves with variable discretizations (number of points and $q$ ranges).

In the following, we first discuss the theoretical concepts necessary to understand our approach in Section II, after which we present the technical details of the implementation in Section III. Finally, we demonstrate the results of our method for different parameterizations of the SLD profile (two-layer box model, five-layer box model, physics-informed model with $N$ repeating monolayer units) on both simulated and experimental reflectivity curves.

\section{Theoretical considerations}

\subsection{The phase problem}

The phase problem is a ubiquitous issue for experimental techniques utilizing the interference of waves  $A e^{-i\omega t}$ as a means of probing the physical properties of materials, caused by the fact that detectors cannot record the phase of the signal, but only its intensity $\left| A \right|^2$ \cite{Volostnikov1990}. This loss of information introduces a degree of ambiguity when trying to reconstruct the physical quantities of interest from the measured signal. It is well-known that neutron and X-ray reflectometry are among the techniques affected by the phase problem\cite{Kozhevnikov_2003}, which precludes an analytical solution via the Gel'fand-Levitan-Marchenko\cite{Newton1974} inverse scattering equation. Some approaches for experimentally tackling the phase problem of neutron and X-ray reflectometry have been developed, such as the reference layer method \cite{Majkrzak1998} and the Lloyd mirage technique \cite{Allman1994}, but they have certain practical limitations. Of particular interest is the use of magnetic reference layers \cite{Masoudi2005} in polarized neutron reflectometry where the SLD of the reference layer depends on the polarization of the incident neutrons. Three distinct measurements for up, down, and non-polarized neutron beams then allow for the determination of the phase. Also for NR, the amount of structural detail extracted from reflectivity data can be maximized by flexibly varying the SLD of specific structures within the sample via isomorphic isotropic substitution \cite{Heinrich2016}, such as the replacement of protium with deuterium (selective deuteration) in the hydrocarbon chains of lipid bilayers \cite{Clifton2012}. Alternatively, the degree of ambiguity can be diminished when measuring a series of reflectivity curves for a sample with evolving structure (e.g. during film deposition). Notably, such theoretical ambiguity is in practice further accentuated by experimental sources of error (noise, finite instrumental resolution, artifacts) and the discrete nature of the measurement process, the scattered intensity being recorded over a finite range of the momentum transfer $q_z$ at a finite number of points.  \\

A consequence of the phase problem is the underdetermined ("ill-posed") nature of the parameter recovery from the measured data, different SLD profiles corresponding to equivalent reflectivity curves. When taking into account multiple scattering at the interfaces, as described by Parratt's recursive formalism \cite{Parratt1954} or Abel{\`{e}}s transfer-matrix method \cite{Abeles1950}, some information about the phase can be recovered in the low $q_z$ region \cite{Zhou1993}, although the difference between reflectivity curves can still become vanishingly small \cite{Pershan1994}, especially when compounded with systematic measurement errors of the total reflection edge. \\

In the following, we briefly summarize a theoretical derivation of how such SLD profiles can be identified in the mathematically simpler kinematical approximation, which neglects multiple scattering at the sample interfaces so that the calculation remains analytical. In the kinematical approximation, the scattered intensity is proportional to the square of the total scattering amplitude:
\begin{equation} \label{eq:kinematical_approximation}
   R(q_z) = \frac{R_\text{F}(q_z)}{\rho_\text{s}^2} \left| \int_{-\infty}^{+\infty} \frac{d\rho(z)}{dz} e^{i q_z z} dz \right|^2 = \frac{R_\text{F}(q_z)}{\rho_\text{s}^2} \left| \mathrm{FT}\left(\frac{d\rho(z)}{dz} \right)\right|^2, 
\end{equation}
where $R_\text{F}(q_z)$ denotes the Fresnel reflectivity, $\rho(z)$ is the SLD profile along the perpendicular direction $z$ to the sample and $\rho_\text{s}$ is the SLD of the substrate.

For the commonly used box model with interfacial roughness parametrized via Névot-Croce factor \cite{Nevot1980}, an SLD profile can be written as a sum of the error functions:
\begin{equation}\label{eq:sld_function}
    \rho(z) = \rho_{\mathrm{ambient}} + \sum_{i=1}^N \Delta\rho_i \cdot \mathrm{erf}\left(\frac{z - z_i}{\sqrt{2}\sigma_i}\right),
\end{equation}
where $\mathrm{erf}(z)$ is the error function, $N$ is the number of interfaces, $z_i$ the position of the $i$th interface, $\sigma_i$ is the roughness of the $i$th interface and $\Delta \rho_i$ is the SLD difference between interfaces $i$ and $i+1$.

By substituting Equation (\ref{eq:sld_function}) into Equation (\ref{eq:kinematical_approximation}), we can explicitly calculate the Fourier transform:

\begin{equation}
    \left| \mathrm{FT}\left(\frac{d\rho(z)}{dz} \right) \right|^2 = \left| \int_{-\infty}^{+\infty} \sum_{i=1}^{N}  \Delta\rho_i \frac{1}{\sqrt{2\pi\sigma^2}} e^{-\frac{(z - z_i)^2}{2\sigma^2}} e^{i q_z z} dz \right|^2 = \left|  \sum_{i=1}^{N}  \Delta\rho_i  e^{iq_zz_i -q_z^2\sigma_i^2/2} dz \right|^2
\end{equation}

Finally, we obtain a decomposition of the scattered intensity into the sum of a constant term and several sinusoidal components with amplitudes $\Delta\rho_i\Delta\rho_j e^{-q_z^2 (\sigma_i^2 + \sigma_j^2) / 2}$ and frequencies $\Delta z_{ij}$:

\begin{align}
    \frac{R_{\text{box}}(q_z)}{R_\text{F}(q_z)} \rho_\text{s}^2 &= \sum_{i=1}^N \Delta\rho_i^2 e^{-q_z^2\sigma_i^2} + 2\sum_{i=2}^N \sum_{j=1}^{i-1} \Delta\rho_i\Delta\rho_j e^{-q_z^2 (\sigma_i^2 + \sigma_j^2) / 2} \cos(q_z \Delta z_{ij}).
\end{align}

\begin{figure}[h]
    \centering
    \includegraphics[width=0.4\linewidth]{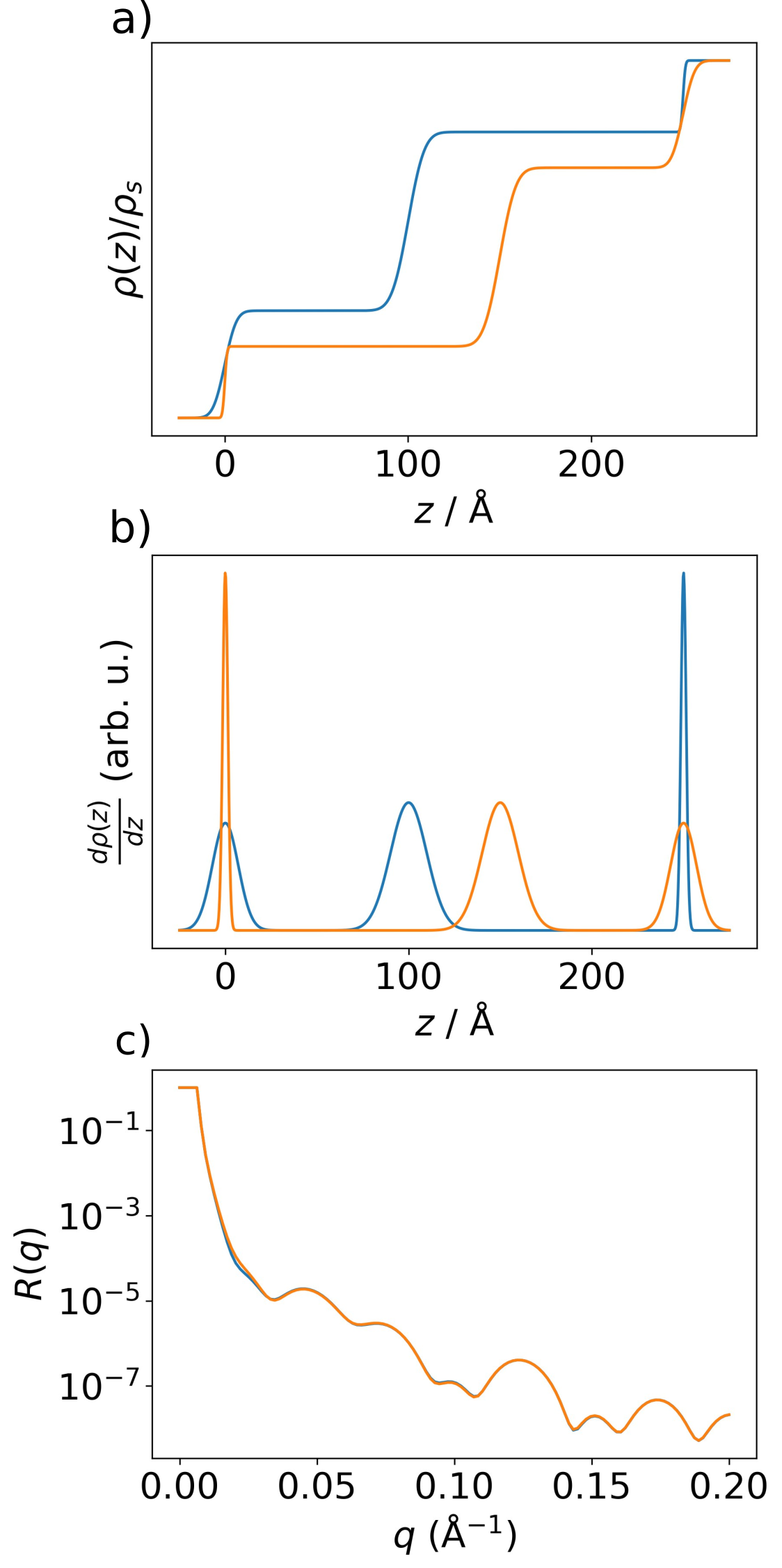}
    \caption{(a) Two SLD profiles with mirrored derivatives (b) The derivatives of the two SLD profiles (c) The corresponding reflectivity curves for the two SLD profiles are almost identical (they would be exactly identical in the kinematical approximation), despite the fact that the SLD profiles in (a) are very dissimilar.}
    \label{fig:mirrored_sld}
\end{figure}

By solving for all combinations of $\Delta\rho_i$ and $z_i$ with the same scattered intensity, one can identify classes of the theoretical identical solutions in the kinematical approximation, such as profiles with mirrored derivatives as reported in earlier studies \cite{Pershan1994, Sivia1991, Pynn1992} (shown in Figure \ref{fig:mirrored_sld}). Notably, restricting the shape of the considered SLD profiles to the box model with a fixed number of layers already imposes a strong regularization on the space of admissible solutions.

\clearpage

\subsection{Solving ill-posed inverse problems using neural networks}
If a forward process $f(\mathbf{x})=\mathbf{y}$ maps the hidden parameters $\mathbf{x}$ describing a physical system to an observable signal $\mathbf{y}$, the inverse problem can be described as retrieving the physical parameters from a possibly noise-corrupted measurement of the signal \cite{Kabanikhin2008}. 
When the inverse problem $f^{-1}(\textbf{y}) = \textbf{x}$ represents a one-to-one mapping, neural networks can be straightforwardly trained to approximate the inverse function. In contrast, a many-to-one mapping of the forward process leads to an underdetermined, ill-posed inverse problem, and the corresponding one-to-many inverse mapping $f^{-1}(\textbf{y})$ cannot be approximated via regression. Attempting to train a neural network as a point estimator over the domain containing the non-uniqueness would lead to incorrect predictions corresponding to an average of distinct solution branches. Different machine learning approaches exist for addressing ill-posed inverse problems \cite{Adler2017, Li_inverse, ardizzone2018analyzing}, which generally depend on the specific application. 

The present study proposes a novel approach for tackling ill-posed inverse problems, which takes advantage of the \textit{a priori} knowledge of the experimentalist. This approach is inspired by conventional reflectometry analysis and combines the high speed of neural networks with the flexibility of conventional fitting procedures.

Typically, some physical parameters of the studied samples are known to the experimentalists with narrow uncertainty ranges, such as densities of the used materials, thicknesses of the deposited layers, etc. In conventional fitting procedures, this information is used to set upper and lower bounds on each sample parameter within which solutions are allowed, the bounds being narrower for layer SLDs and wider for layer thicknesses and roughnesses. Our implementation allows us to provide such prior knowledge as input to the neural network in the form of the two prior bounds for each parameter, the regression problem being constrained to the local domain defined by the prior bounds, thus avoiding the non-uniqueness associated with the full parameter domain. As long as the prior bounds are narrow enough not to include multiple solution branches, the inverse mapping is well-determined and can be approximated by a neural network. In practice, even a small amount of prior knowledge can be enough to rule out ambiguity in reflectometry analysis. We note that narrow priors do not guarantee a single solution, as multiple solutions might occur close to each other. However, our general assumption shared by experimentalists who analyze reflectometry data via conventional fitting is that for a large set of cases, the prior information is sufficient to isolate a single solution. 
 
 Our method can be envisioned as simultaneously learning the inverse problem on all the possible subdomains (or a subset of subdomains) of the full parameter domain, using a single neural network. Thus, some parallels can be drawn to approaches such as \cite{bae2022multirate} where a Beta Variational Autoencoder ($\beta$-VAE) is trained simultaneously for all values of the $\beta$ coefficient using a single neural network (a task which would typically require retraining for each $\beta$ value).

\subsection{Discretization-invariant learning}

Neural networks can only learn mappings between finite-dimensional vector spaces, some architectures such as the multilayer perceptron requiring a fixed discretization (range and resolution) of the input. A new paradigm is represented by neural operators, which can learn mappings between infinite-dimensional function spaces, allowing discretization-invariant learning \cite{Li2020, kovachki2023neural}.  For an input function $v_{0}(x)$ defined over a domain $D$, a neural operator is constructed as a series of transformations $v_t \mapsto v_{t+1}$:
$$v_{t+1}(x) = \sigma \left (W v_{t}(x) 
+ (\mathfrak{K}v_{t})(x)  \right )$$
where $W$ is a linear transformation, $\sigma$ is a non-linear activation function and $\mathfrak{K}$ is a non-local integral operator with learnable kernel $\mathfrak{k}$:
$$ (\mathfrak{K}v_{t})(x) = \int_{y\in D} \mathfrak{k}(x,y) v_{t}(y) dy $$
The Fourier Neural Operator (FNO) \cite{li2021fourier} represents an efficient and expressive implementation of a neural operator which imposes $\mathfrak{k}(x,y) = \mathfrak{k}(x-y)$ and leverages the convolution theorem. It parameterizes the kernel operator directly in Fourier space as:
$$ (\mathfrak{K}v_{t})(x) = \mathfrak{F}^{-1}\left ( T\left ( \mathfrak{F} v_{t}  \right ) \right )(x) $$
where $\mathfrak{F}$ and $\mathfrak{F}^{-1}$ represent the direct and, respectively, the inverse discrete Fourier transform and $T$ is a learnable linear transformation.

Neural operators have been predominantly used in the fields of differential equation solving and physics-informed learning \cite{Oommen2022, Wen2022}. Here, we use a neural operator to learn a vector embedding for reflectivity curves with variable discretizations ($q$ ranges and numbers of points) for our regression inverse problem. Such an approach is beneficial as it confers a higher degree of flexibility to the trained model, enabling the use of the full measured signal without relying on interpolation.

\section{Design and implementation of our method}

\subsection{Training methodology and neural network architecture}

To enable the regularization of the inverse problem by confining the regression within a sample-dependent local domain, we generate ground truth values of the parameters for training in the following manner. Firstly, for each parameter, we obtain the center ($c$) and the width ($w$) of the local domain, the center being uniformly sampled from the global domain (i.e. the parameter range) and the width being uniformly sampled from a predefined width range for that parameter. Secondly, the ground truth values of the parameters are obtained by uniform sampling within the local domain $[c-w/2, c+w/2]$ defined by the previously sampled values, the two prior bounds being the edges of this local domain, $c-w/2$ and $c+w/2$. 

At each training step, we simulate a new batch of reflectivity curves from the sampled ground truth parameters using a fast, GPU-accelerated Pytorch \cite{Paszke2019} implementation of the Abel{\`{e}}s transfer-matrix method. As a preprocessing step, we set intensities below 10$^{-10}$ which cannot be recorded in most experimental scenarios (although there are exceptions) to this chosen minimum threshold, we add noise to the curve, and we apply a logarithmic transformation followed by a linear rescaling. The prior bounds are normalized with respect to the corresponding parameter ranges and the ground truth parameters are normalized with respect to the local domain they were sampled from, such that all the inputs and outputs to the neural network are in the range [$-$1, 1].

\begin{figure}[h]
    \centering
    \includegraphics[width=1.0\linewidth]{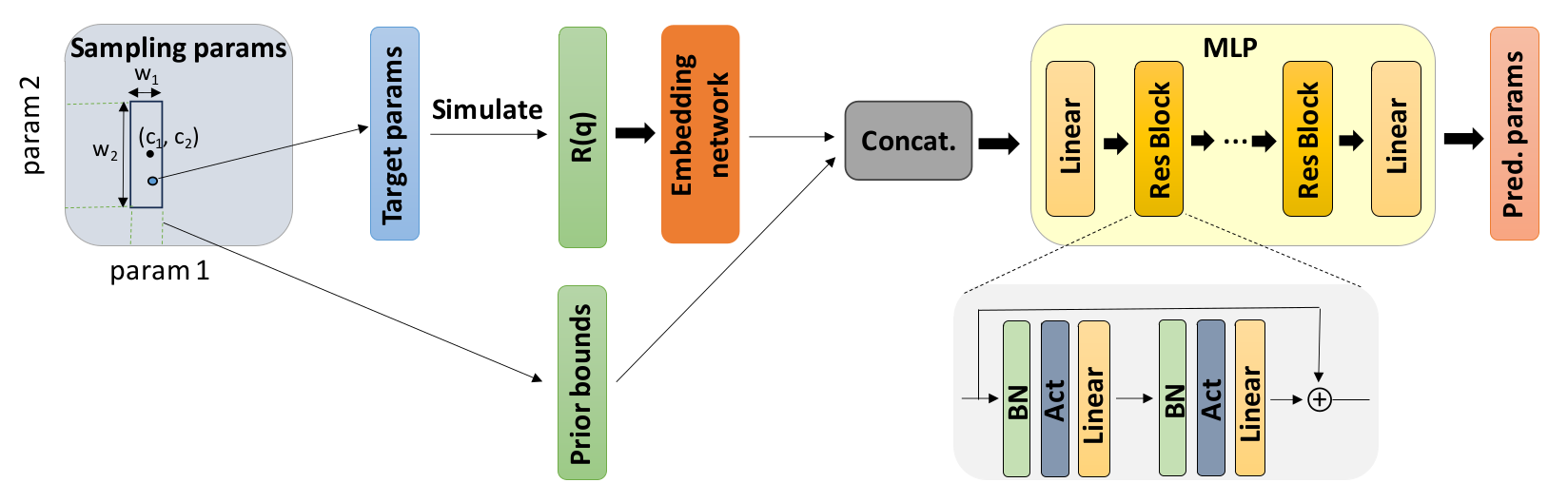}
    \caption{Neural network architecture and steps involved in our method. Firstly, the ground truth parameters and the prior bounds are sampled using the described procedure. An embedding of the reflectivity curves together with the prior bounds are provided as input to the MLP. The MLP consists of residual blocks with batch normalization (BN), nonlinear activation (chosen to be GELU) and linear layers.}
    \label{fig:network_drawing_mlp}
\end{figure}

As shown in Figure \ref{fig:network_drawing_mlp}, an embedding of the reflectivity curve, together with the prior bounds are input to a fully connected neural network (also known as multilayer perceptron or MLP \cite{Murtagh1991}), the loss being computed as the mean-squared error between the neural network output and the ground truth parameters (normalized with respect to the prior bounds). Notably, obtaining the final prediction requires reversing the normalization of the neural network output with respect to the prior bounds. 
The architecture of our model, as well as the sampling procedure, are shown in Figure \ref{fig:network_drawing_mlp} (parameter scaling omitted for simplicity). The MLP consists of a sequence of $n_\text{blocks}$=6 residual blocks inspired by the ResNet\cite{He2016} architecture design, each block containing 2 hidden layers of width dim$_\text{hidden}$=1024 neurons. The use of skip connections has the role of facilitating gradient propagation and preventing singularities \cite{orhan2018skip}. Batch normalization \cite{pmlr-v37-ioffe15} is known to improve the convergence of neural networks, so we use it to normalize the intermediate features before activation. Since the type of activation function can have a significant impact on the performance of a neural network, we explored the use of several popular activation functions (ReLU\cite{Fukushima1975}, GELU\cite{hendrycks2020gaussian}, SELU\cite{Klambauer2017}, Mish\cite{misra2020mish}), choosing GELU as the default.

We trained our models using the AdamW\cite{loshchilov2018decoupled} optimizer, a version of Adam\cite{kingma2017adam} with decoupled weight decay regularization. The weight decay coefficient is kept at the default value of 0.01 . The initial learning rate of 0.0001 was decreased by a factor of 10 on plateau of the loss. We used the largest batch size that fits in our GPU memory (4096) to ensure stable gradients. 

\subsection{Embedding networks}

Going beyond previous publications, where a reflectivity curve is directly provided as input to the MLP, we first use a network that produces a latent embedding of the reflectivity curve which is subsequently fed to the MLP.

\begin{figure}[h]
    \centering
    \includegraphics[width=0.6\linewidth]{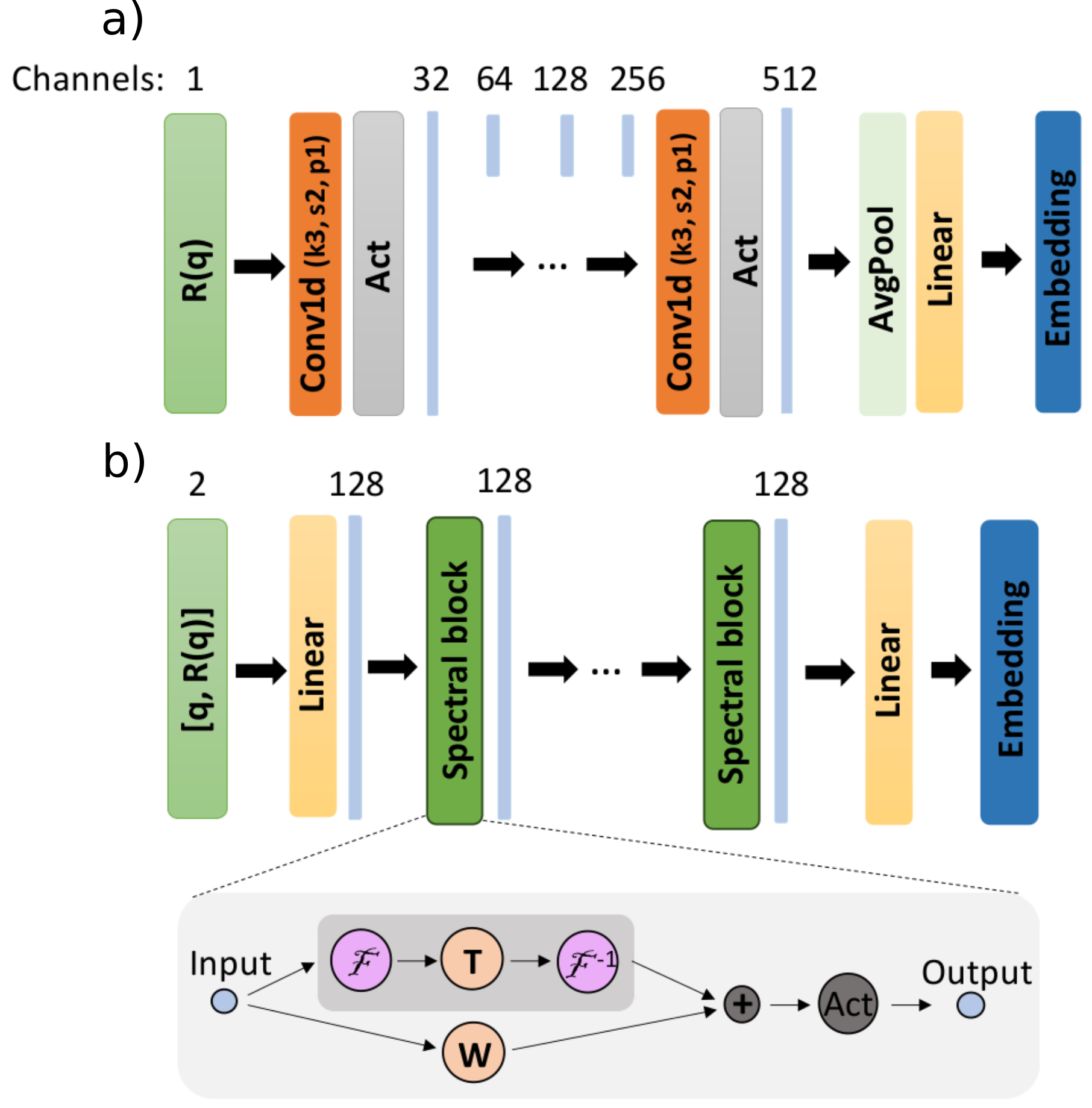}
    \caption{Architecture of the embedding networks (a) 1D CNN, consisting of convolutions with kernel size 3, stride 2 and padding 1 (b) FNO, consisting of spectral blocks which implement the neural operator kernel in Fourier space.}
    \label{fig:embedding_networks}
\end{figure}

When training a model on reflectivity curves with fixed discretization, we use a 1D CNN \cite{Kiranyaz2021} as the embedding network, which is parameter-efficient and better leverages the sequential characteristics of the data. While less popular than their 2D counterparts, 1D CNNs have been successfully used in a variety of tasks involving 1D signals such as automatic speech recognition \cite{Collobert2016} and time-series prediction \cite{Guessoum2022}. The 1D CNN, as shown in Figure \ref{fig:embedding_networks}a, consists of a sequence of convolutional layers with kernel size 3, stride 2 and padding 1, the dimension of the signal being (approximately) halved after each layer. At the same time, the number of channels is doubled after each layer starting from 32 up to ch$_\text{out}$=512. An adaptive average pooling layer with output size dim$_\text{avpool}$=8 ensures a fixed input size (ch$_\text{out}$*dim$_\text{avpool}$) for a linear layer which produces the final embedding with dimension dim$_\text{emb, CNN}$ = 128. While the adaptive average pooling allows obtaining a fixed size embedding from curves with variable discretizations, CNNs do not enable discretization-invariant learning, as shown in previous studies \cite{li2021fourier}. 

When training a model on reflectivity curves with variable discretizations, we employ a Fourier Neural Operator (FNO) as the embedding network (\ref{fig:embedding_networks}b), as theoretically motivated in the previous section on discretization-invariant learning. In this scenario the minimum and maximum values of $q$, as well as the number of points in the curve are also uniformly sampled for each batch from the considered ranges. The input to the FNO is the reflectivity curve together with the corresponding $q$ values (concatenated along the channel axis). After the input is raised to a higher channel space ch$_\text{FNO}$=128 by a pointwise linear operation, a sequence of $n_\text{FNO}$=5 spectral blocks are applied which implement the kernel operator in Fourier space as illustrated in Figure \ref{fig:embedding_networks}b. The number of Fourier modes kept after performing the discrete Fourier transform is a hyperparameter set at $n_\text{modes}$=16. Finally, a mean pooling over the input dimension followed by a linear layer produces the final embedding with dimension dim$_\text{emb, FNO}$ = 256.

\newpage

\section{Results}

We trained neural networks according to the previously described approach for different parameterizations of the thin film SLD profile. Each of the following subsections elaborates on a network trained for data with a different parameterization: either a different number of layers (two and five) for the box model or a physics-informed special parameterization of a multilayer structure with repeating monolayers. We use the 1D CNN as the default embedding network. In the last subsection the FNO is used as the embedding network for data with variable discretization. We evaluate performance metrics on statistically significant batches of simulated data and we illustrate the applicability of our method on experimental reflectivity data when available.

\clearpage

\subsection{Two-layer box model}

This subsection shows the performance evaluation for a neural network trained on data corresponding to the two-layer parameterization of the box model, the total number of physical parameters predicted by the neural network being 8 (3 parameters per layer (thickness, roughness, SLD) plus 2 additional parameters for the substrate (roughness and SLD)). The $q$ range of the simulated curves is [0.02, 0.15] \AA$^{-1}$ which aligns with the range of the test experimental data, and the resolution is 128 points. 
Various types of noise were applied to the simulated reflectivity curves in order to provide robustness to experimental artifacts, namely Poisson noise, $q$-position noise, curve shifting and curve scaling based on previous investigations \cite{Greco2021Neural}.

\begin{figure}[h]
    \centering
    \includegraphics[width=1.0\linewidth]{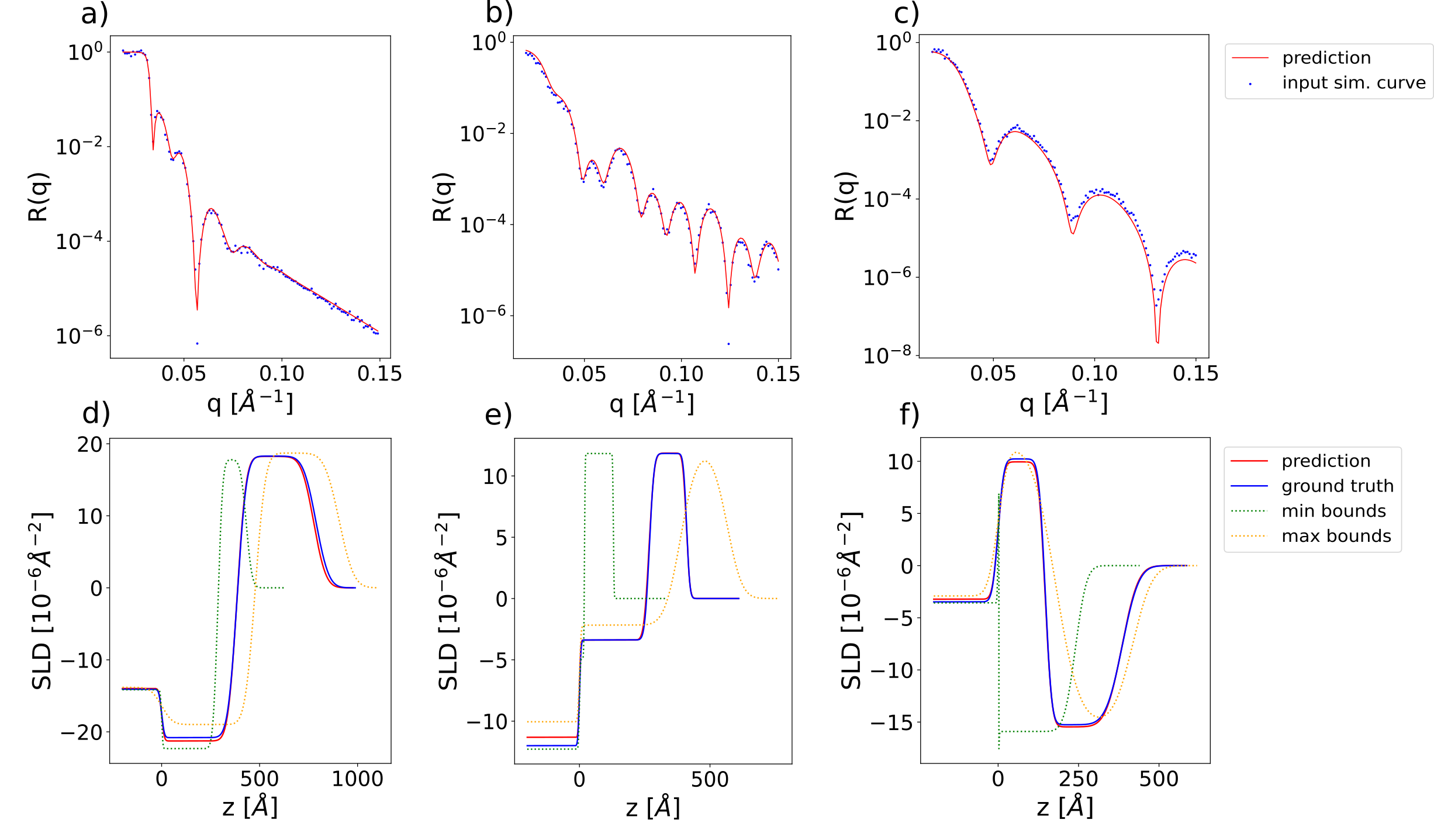}
    \caption{(a)-(c) Examples of input simulated reflectivity curves (blue markers) and the corresponding neural network predictions (red line) for the 2-layer model. (d)-(f) Ground truth (blue) and predicted (red) SLD profiles corresponding to the reflectivity curves in the top row. SLD profiles corresponding to the minimum (green) and maximum (orange) prior bounds used for the prediction are also shown as dotted lines.}
    \label{fig:example_refl_curves_2layers}
\end{figure}

\begin{figure}[h]
    \centering
    \includegraphics[width=0.8\linewidth]{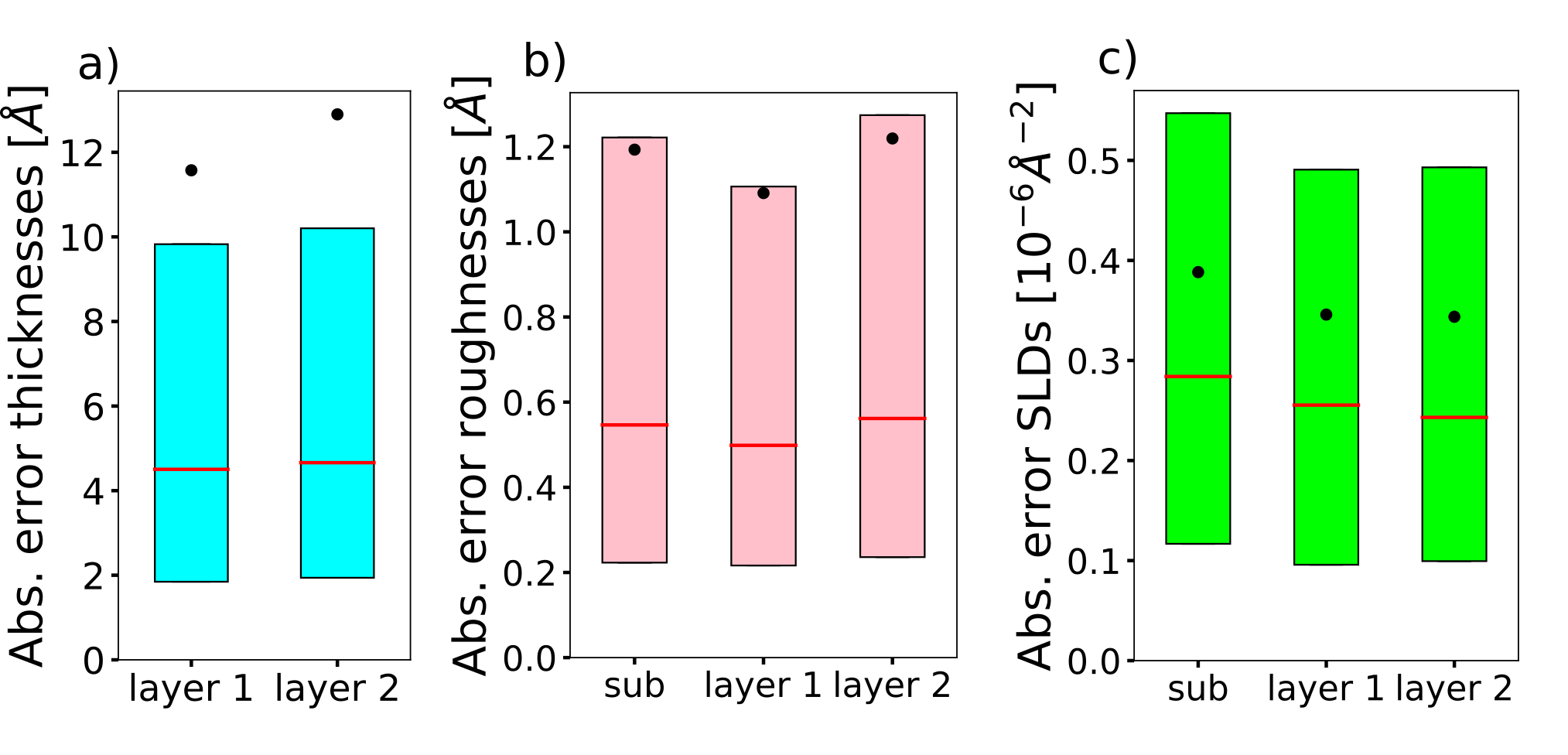}
    \caption{Boxplots of the absolute errors of each predicted parameter of the 2-layer model, computed over a batch of 4096 simulated curves, with the prior bounds being uniformly sampled. The horizontal red line denotes the median, the black dot denotes the mean and the lower and upper extremities of the box plot denote the 25$^\text{th}$ percentile and the 75$^\text{th}$ percentile respectively.}
    \label{fig:results_sim_2layers_boxplots}
\end{figure}

 The chosen parameter ranges are [0, 500] {\AA} for the thicknesses, [0, 60] {\AA} for the roughnesses, and [$-$25, 25] $10^{-6} ${\AA}$^{-2}$ for the SLDs. Notably, the negative values of the SLD include specific cases of NR. The ranges of the prior bound widths are [0.01, 500] {\AA} for the thicknesses, [0.01, 60] {\AA} for the roughnesses and [0.01, 4] $10^{-6} ${\AA}$^{-2}$ for the SLDs. While the prior bound widths for thicknesses and roughnesses span the whole domain of these parameter types, the maximum prior bound width for the SLDs was reduced, since this type of parameter is associated with the highest amount of prior experimental knowledge. Figure \ref{fig:example_refl_curves_2layers} illustrates examples of input simulated reflectivity curves for this model together with the neural network predictions (top row) as well as the ground truth and predicted SLD profiles (bottom row). Figure \ref{fig:results_sim_2layers_boxplots} shows the box plots of the absolute errors for each of the 8 predicted parameters ($a.$ thicknesses, $b.$ roughnesses, $c.$ SLDs), computed over a batch of 4096 simulated curves, the ground truth parameters and prior bounds being generated as in the training procedure. We can see from the box plots that the prediction errors for thicknesses, roughnesses and SLDs are quite low considering the large parameter ranges used for training. 
 
\begin{figure}[h]
    \centering
    \includegraphics[width=1.0\linewidth]{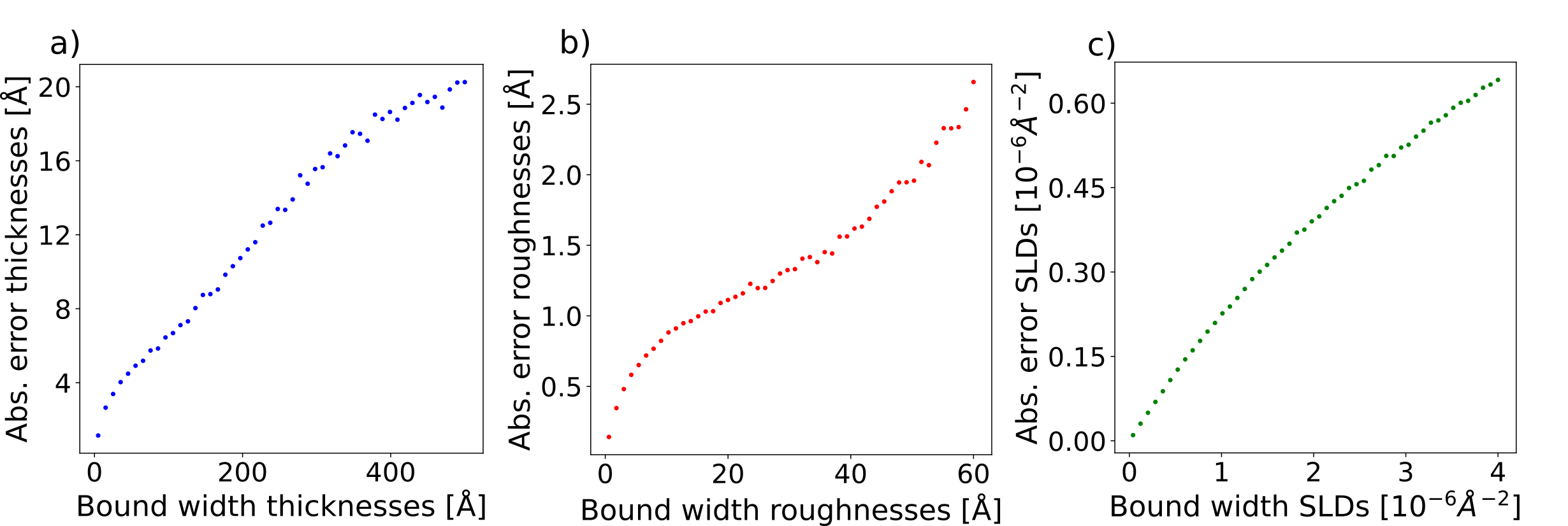}
    \caption{Dependence of the mean absolute error of each parameter type ($a.$ thickness, $b.$ roughness, $c.$ SLD) for the 2-layer model as a function of the prior bound width.}
    \label{fig:results_sim_2layers_bound_width_dependence}
\end{figure}

 It is important to understand how the performance of the model depends on the input prior bounds, as this informs users how much prior knowledge they should provide for an expected prediction quality. To evaluate such a dependence, we sample batches of ground truth parameters and prior bounds such that the prior bound widths are fixed for each batch. The prior bound widths are varied by multiplying the maximum bound width of one parameter type at a time (while for the other parameter types keeping the maximum bound width) by a scalar value in the range [0, 1]. As shown in Figure \ref{fig:results_sim_2layers_bound_width_dependence}, we observe that the absolute errors of the thicknesses, roughnesses and SLDs decrease when the relative prior bound width for the corresponding parameter type is decreased, as expected. A relative bound width of 0 represents the trivial case when the prior bounds define the ground truth parameters exactly. 
 
\begin{figure}[h]
    \centering
    \includegraphics[width=1.0\linewidth]{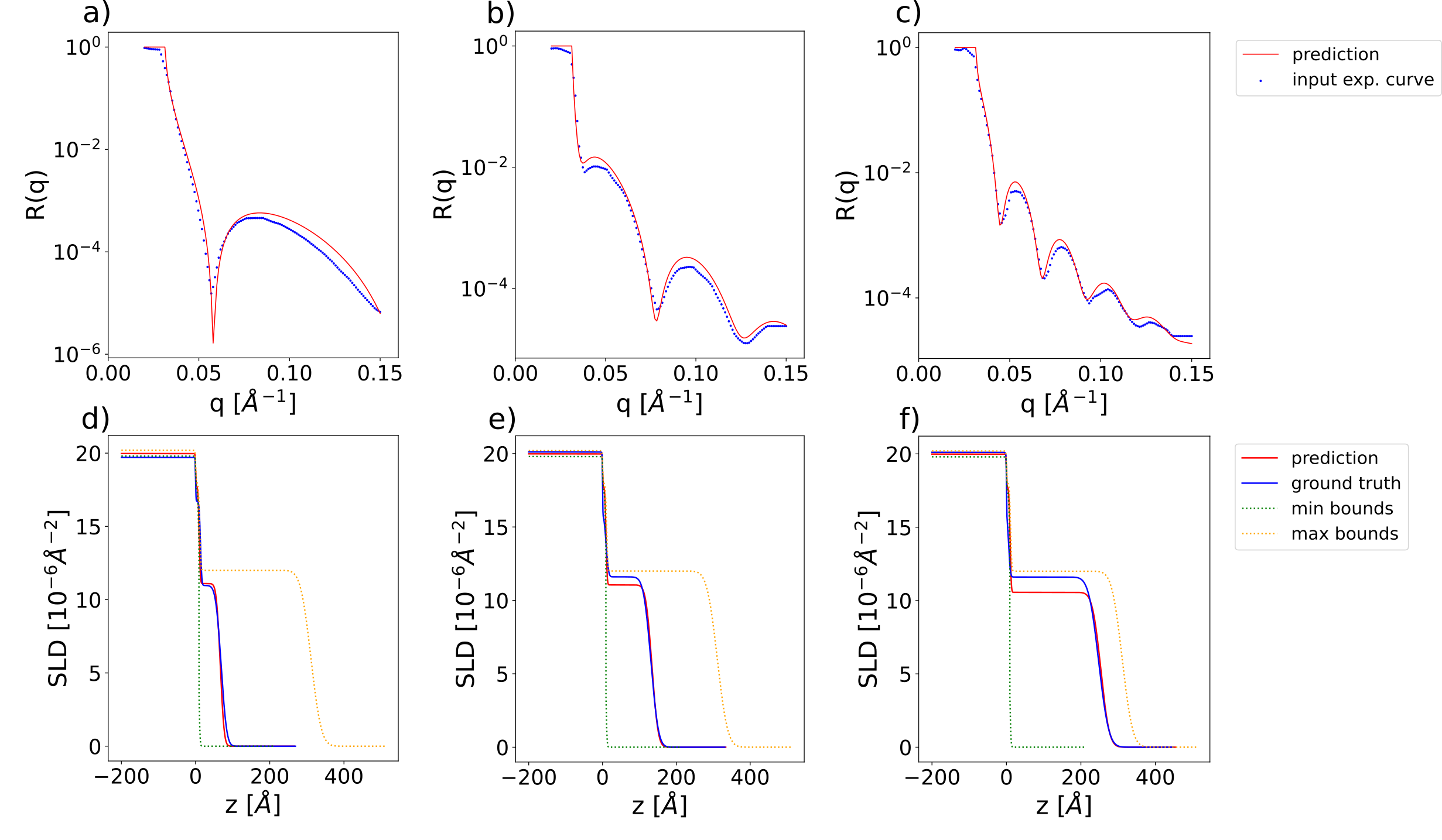}
    \caption{(a)-(c) Examples of input experimental reflectivity curves (blue markers) and the corresponding neural network predictions (red line). (d)-(f) Ground truth (blue) and predicted (red) SLD profiles corresponding to the reflectivity curves in the top row. SLD profiles corresponding to the minimum (green) and maximum (orange) prior bounds used for the prediction are also shown as dotted lines.}
    \label{fig:example_refl_curves_exp}
\end{figure}

We demonstrate the applicability of our method for analyzing experimental reflectivity data using XRR curves from a previously published dataset \cite{Pithan2022}, containing \textit{in situ} measurements performed during the deposition of a layer of organic material of diindenoperylene (DIP) or pentacene (PEN), on top of a thin silicon oxide layer sitting on a silicon substrate, together with the ground truth manual fits of the parameters. We are able to leverage the \textit{a priori} experimental knowledge about this dataset by setting narrow prior bounds around the known values of the SLD for the substrate (Si) and the bottom layer (SiO$_2$). In previous approaches \cite{Greco2022Neural} using this dataset, all the parameters of the substrate and SiO$_2$ layer were kept at predefined constant values during training and only the thickness, roughness and SLD of the top layer were predicted. Our approach is capable of also tackling this specific use case while being trained for more general cases. Figure \ref{fig:example_refl_curves_exp} shows input experimental curves together with the neural network predictions (top row), as well as the corresponding SLD profiles (bottom row).

\clearpage

\subsection{Five-layer box model}

\begin{figure}[h]
    \centering
    \includegraphics[width=1.0\linewidth]{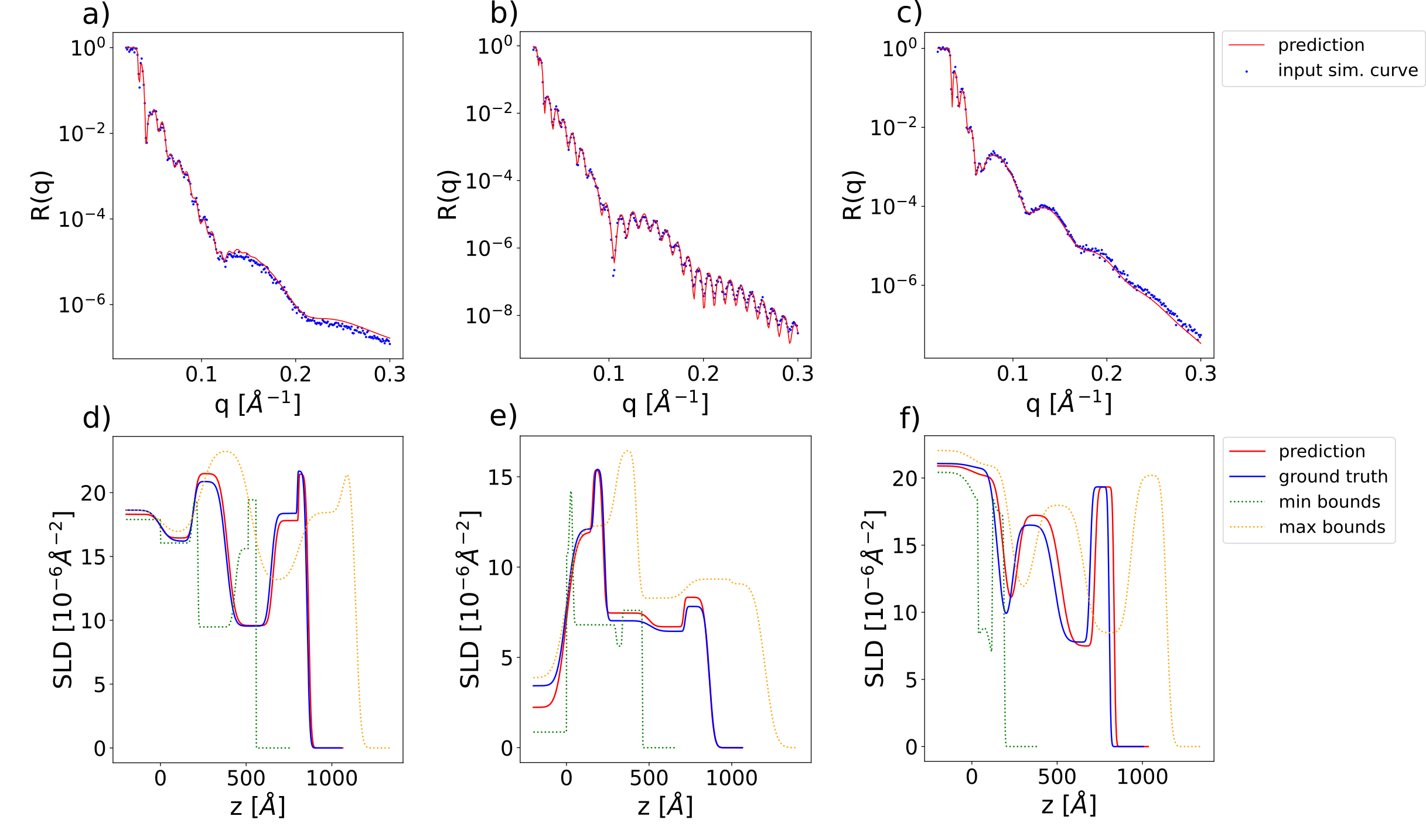}
    \caption{(a)-(c) Examples of input simulated reflectivity curves (blue markers) and the corresponding neural network predictions (red line) for the 5-layer model. (d)-(f) Ground truth (blue) and predicted (red) SLD profiles corresponding to the reflectivity curves in the top row. SLD profiles corresponding to the minimum (green) and maximum (orange) prior bounds used for the prediction are also shown as dotted lines.}
    \label{fig:example_refl_curves_5layers}
\end{figure}

\begin{figure}[h]
    \centering
    \includegraphics[width=1.0\linewidth]{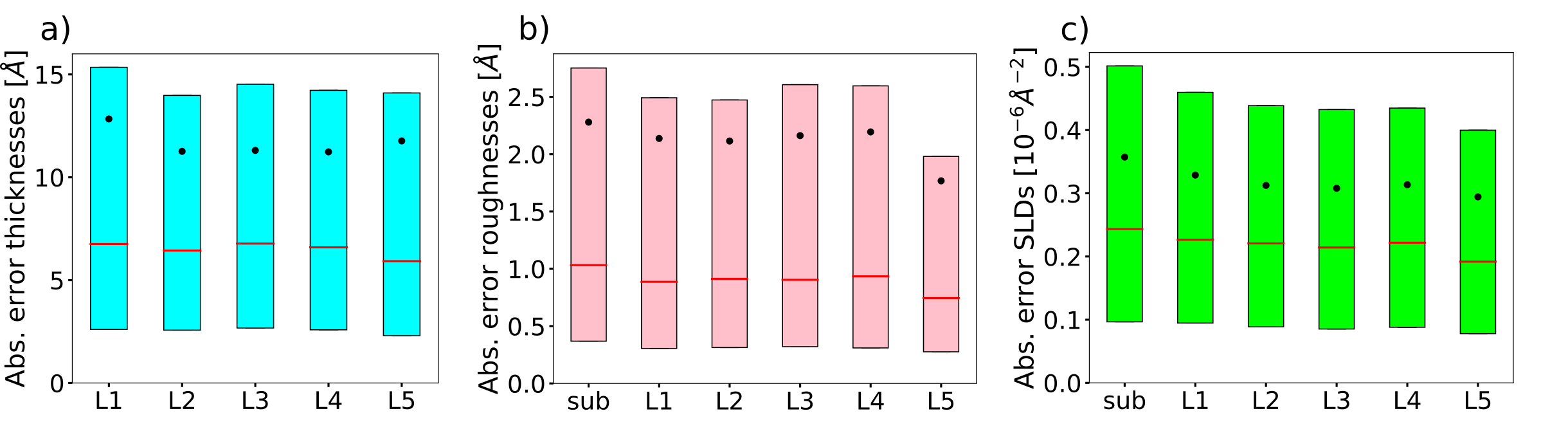}
    \caption{Boxplots of the absolute errors of each predicted parameter of the 5-layer model, computed over a batch of 4096 simulated curves, with the prior bounds being uniformly sampled. The horizontal red line denotes the median, the black dot denotes the mean and the lower and upper extremities of the box plot denote the 25$^\text{th}$ percentile and the 75$^\text{th}$ percentile respectively.}
    \label{fig:results_sim_5layers_boxplots}
\end{figure}

Increasing the number of layers in the box model parameterization increases the difficulty of training a neural network for the inverse problem since the reflectivity curves become more complex and also the degree of non-uniqueness is increased. Nevertheless, we demonstrate that our method can still be successfully applied to models with an increased number of layers, namely a five-layer model, the total number of physical parameters predicted by the neural network being 17 (3 parameters per layer (thickness, roughness, SLD) plus 2 additional parameters for the substrate (roughness and SLD)). We increase the $q$ range of the simulated curves to [0.02, 0.3] \AA$^{-1}$ and the resolution to 256 points. The chosen parameter ranges are [0, 300] {\AA} for the thicknesses, [0, 60] {\AA} for the roughnesses, and [0, 25] $10^{-6} ${\AA}$^{-2}$ for the SLDs. The ranges of the prior bound widths are [0.01, 300] {\AA} for the thicknesses, [0.01, 60] {\AA} for the roughnesses, and [0.01, 4] $10^{-6} ${\AA}$^{-2}$ for the SLDs. Figure \ref{fig:example_refl_curves_5layers} shows examples of input simulated curves and predictions for the 5-layer model, together with the corresponding SLD profiles. The absolute errors between the ground truth and the predicted parameters are displayed in Figure \ref{fig:results_sim_5layers_boxplots}. We observe that the prediction errors do not vary much depending on the specific layer in the model. While, as expected, the prediction errors for the 5-layer model are higher than the prediction errors for the 2-layer model (the difference being more pronounced for roughnesses and less pronounced for SLDs), the performance of our model is still very good on this more challenging use case.

\clearpage

\subsection{Complex multilayer model}
\begin{figure}[h]
    \centering
    \includegraphics[width=0.7\linewidth]{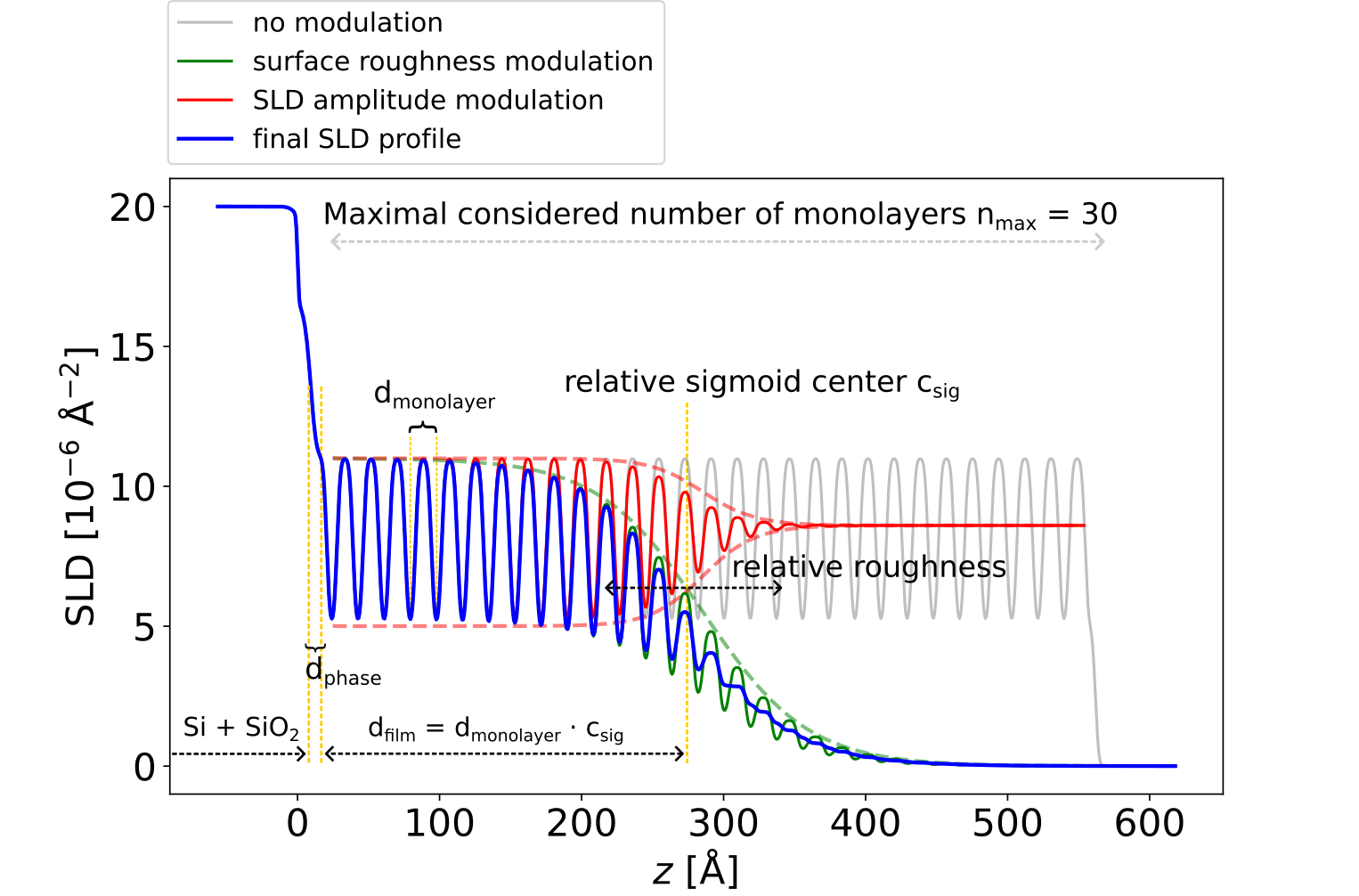}
    \caption{Physics-informed parameterization of the SLD profile for a thin film consisting of repeating identical monolayers on top of a substrate. The grey curve shows the base SLD profile of the monolayers, the green curve shows the SLD profile with surface roughness and the red curve shows the modulation of the SLD amplitude. The blue curve represents the final SLD profile.}
    \label{fig:sketch_multilayer}
\end{figure}

\begin{figure}[h]
    \centering
    \includegraphics[width=0.9\linewidth]{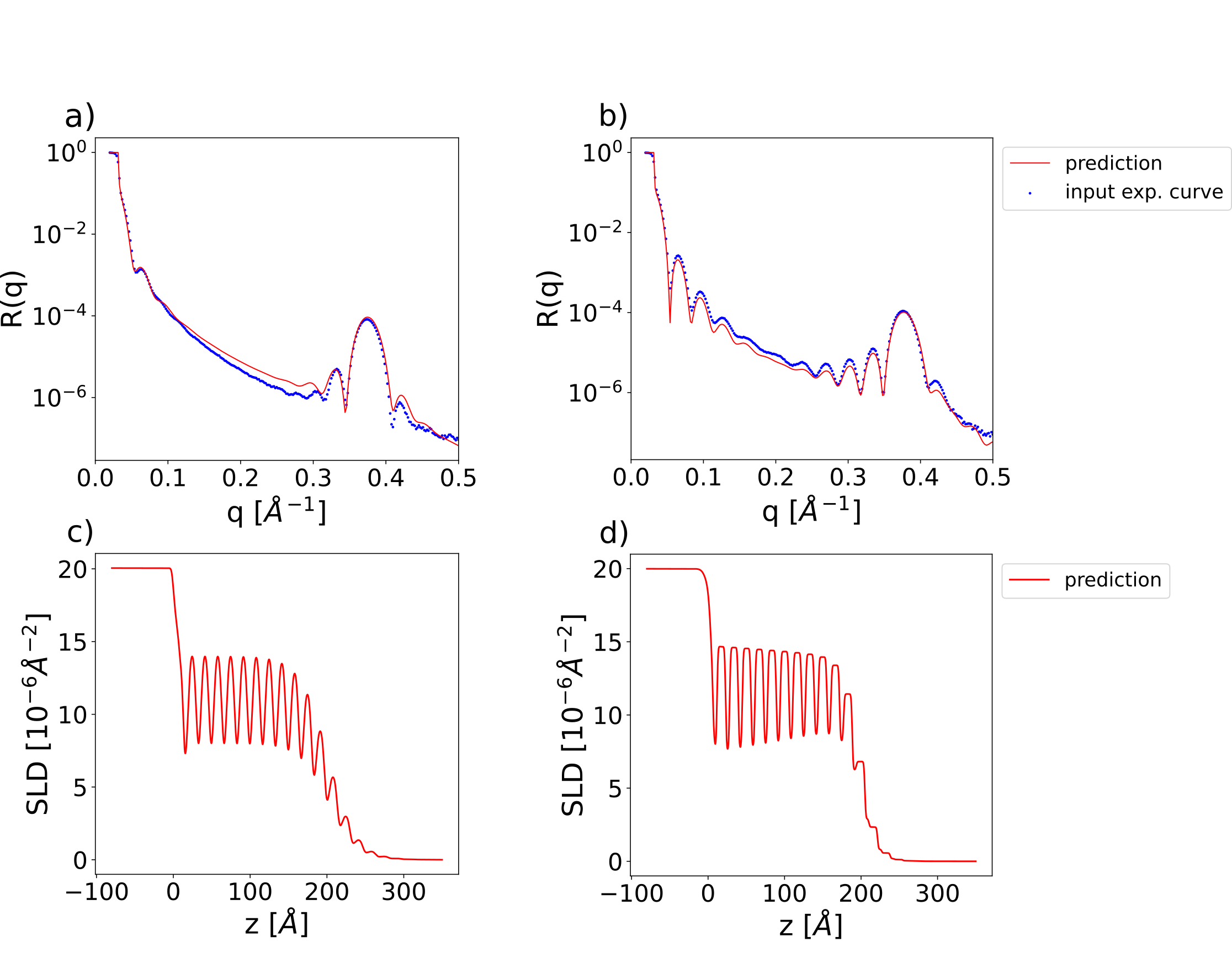}
    \caption{(a)-(b) Examples of input experimental reflectivity curves of DIP monolayers grown on top of a silicon/silicon oxide substrate (blue markers) with corresponding neural network predictions (red lines) for the model with physics-informed parameterization. (c)-(d) Predicted SLD profiles corresponding to the reflectivity curves in the top row.}
    \label{fig:example_refl_curves_multilayer}
\end{figure}

In this subsection, instead of the box model parameterization, we employ a physics-informed parameterization of a complex multilayer structure, as illustrated in Figure \ref{fig:sketch_multilayer}. The physical scenario is the following: on top of a silicon/silicon oxide substrate we consider a thin film composed of repeating identical monolayers (grey curve in Figure \ref{fig:sketch_multilayer}), each monolayer consisting of two boxes with distinct SLDs. A sigmoid envelope modulating the SLD profile of the monolayers defines the film thickness and the roughness at the top interface (green curve in Figure \ref{fig:sketch_multilayer}). A second sigmoid envelope can be used to modulate the amplitude of the monolayer SLDs as a function of the displacement from the position of the first sigmoid (red curve in Figure \ref{fig:sketch_multilayer}). These two sigmoids allow one to model a thin film that is coherently ordered up to a certain coherent thickness and gets incoherently ordered or amorphous toward the top of the film. Such a scenario is sometimes encountered when Kiessig and Laue fringes show different periods.  In addition, a layer between the substrate and the multilayer is introduced to account for the interface structure, which does not necessarily have to be identical to the multilayer period. This so-called "phase layer" (i.e. a layer that strongly influences the relative scattering phase between the substrate and the multilayer) is important as the relative phase between strong substrate reflection and a multilayer Bragg reflection can lead to very different shapes of the curve around the Bragg reflection for constructive or destructive interference. The 17 parameters describing the model together with their training ranges are displayed in Table 1. For experimental XRR curves of DIP monolayers, measured at a laboratory X-ray source, we leverage prior knowledge about this system (layer spacing, substrate SLDs, approximate SLD values for the two boxes in the monolayer) to set suitable prior bounds. Figure \ref{fig:example_refl_curves_multilayer} shows input experimental curves together with the neural network predictions for the model with physics-informed parameterization. Again the prediction quality is good, demonstrating the potential of the proposed method to fit experimental data in increasingly complex scenarios such as multilayer structures featuring Bragg peaks. It is worth noting that in this case the gradient descent polishing procedure introduced in previous studies \cite{Greco2022Neural} can further improve the fit and it can be performed both with the 17 introduced parameters and with the full set of box model parameters, allowing the final solution to potentially evolve beyond the designed parameterization.

\begin{table}
\centering
\caption{Parameters of the model with physics-informed parameterization of the SLD profile, together with the parameter ranges and the ranges of the prior bound widths used for training. Some of the parameters are relative with respect to the monolayer thickness.}
\begin{tblr}{
  width = \linewidth,
  colspec = {Q[463]Q[267]Q[212]},
  cells = {c},
  hlines,
  vlines,
}
\textbf{Parameter}                                                          & \textbf{Parameter range}                 & \textbf{Prior bound width range}            \\
monolayer thickness                                                         & {[}10, 20] \AA                           & {[}0.1 , 10]~\AA                            \\
relative roughness of the monolayer interfaces                              & {[}0, 0.3]                               & {[}0.1, 0.3]                                \\
SLD of the first box in the monolayer                                       & {[}0, 20]~$10^{-6} \text{\AA}^{-2}$      & {[}0.1, 5]~$10^{-6} \text{\AA}^{-2}$        \\
SLD difference between the second and the first box in the monolayer        & {[}-10, 10]~$10^{-6} \text{\AA}^{-2}$    & {[}0.1, 5]~$10^{-6} \text{\AA}^{-2}$        \\
fraction of the monolayer thickness belonging to the first box              & {[}0.01, 0.99]                           & {[}0.01, 1]                                 \\
roughness of the silicon substrate                                          & {[}0, 10] \AA                            & {[}0.01, 10] \AA                            \\
SLD of the silicon substrate                                                & {[}19, 21]~$10^{-6} \text{\AA}^{-2}$     & {[}0.01, 2]~$10^{-6} \text{\AA}^{-2}$       \\
thickness of the silicon oxide layer                                        & {[}0, 10] \AA                            & {[}0.01, 10] \AA                            \\
roughness of the silicon oxide layer                                        & {[}0, 10] \AA                            & {[}0.01, 10] \AA                            \\
SLD of the silicon oxide layer                                              & {[}17, 19]~$10^{-6} \text{\AA}^{-2}$     & {[}0.01, 2]~$10^{-6} \text{\AA}^{-2}$       \\
SLD of the phase layer                                                      & {[}0, 25]~$10^{-6} \text{\AA}^{-2}$      & {[}0.01, 25]~$10^{-6} \text{\AA}^{-2}$      \\
relative thickness of the phase layer                                       & {[}0, 1]                                 & {[}0.01, 1]                                 \\
relative roughness of the phase layer                                       & {[}0, 1]                                 & {[}0.01, 1]                                 \\
relative position of the first sigmoid (total film thickness)               & {[}0, 25]                                & {[}0.1, 25]                                 \\
relative width of the first sigmoid                                         & {[}0, 5]                                 & {[}0.1, 5]                                  \\
relative position of the second sigmoid (coherently ordered film thickness) & {[}-10, 10]                              & {[}0.1, 20]                                 \\
relative width of the second sigmoid                                        & {[}0, 20]                                & {[}0.1, 20]                                 
\end{tblr}
\end{table}

\clearpage

\subsection{Model with Fourier Neural Operator embedding network}

\begin{figure}[h]
    \centering
    \includegraphics[width=1.0\linewidth]{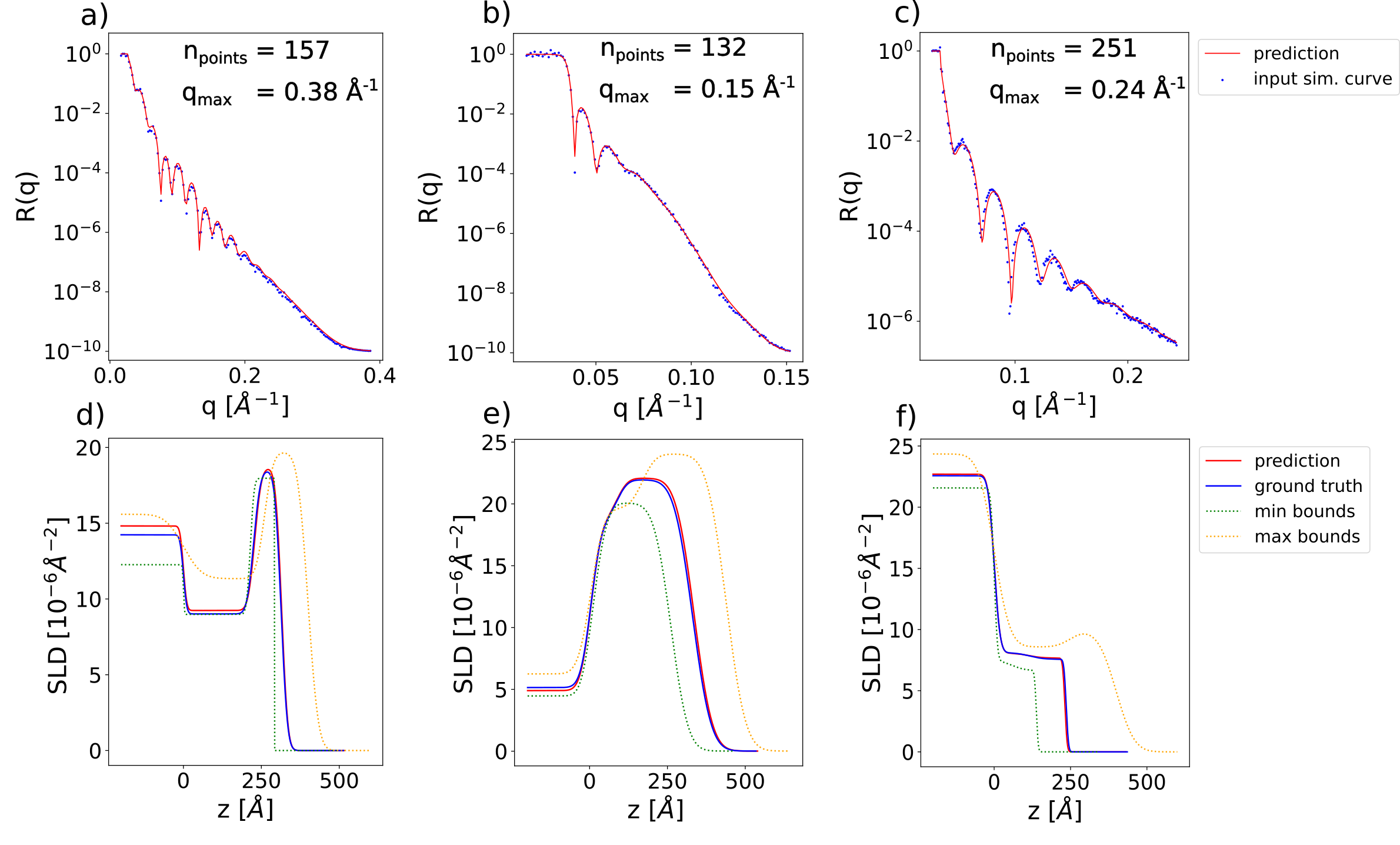}
    \caption{(a)-(c) Examples of input simulated reflectivity curves with variable discretizations (blue markers) and the corresponding neural network predictions (red line) for a 2-layer model with FNO embedding network. (d)-(f) Ground truth (blue) and predicted (red) SLD profiles corresponding to the reflectivity curves in the top row. SLD profiles corresponding to the minimum (green) and maximum (orange) prior bounds used for the prediction are also shown as dotted lines.}
    \label{fig:example_refl_curves_sim_fno}
\end{figure}

In this subsection, we show results obtained when using the FNO as the embedding network instead of the 1D CNN used in the previous sections. The number of points in the simulated curves are in the range [128, 256], the minimum value of $q$ is in the range [0.01, 0.03] {\AA}$^{-1}$ and the maximum value of $q$ is in the range [0.15, 0.4] {\AA}$^{-1}$. We consider a 2-layer box model with parameter ranges [0, 300] {\AA} for the thicknesses, [0, 60] {\AA} for the roughnesses, and [0, 25] $10^{-6} ${\AA}$^{-2}$ for the SLDs. The ranges of the prior bound widths are [0.01, 300] {\AA} for the thicknesses, [0.01, 60] {\AA} for the roughnesses and [0.01, 4] $10^{-6} ${\AA}$^{-2}$ for the SLDs. Due to increased memory demands the batch size is decreased to 1024. Figure \ref{fig:example_refl_curves_sim_fno} shows input curves with variable number of points and $q$ ranges, together with the neural network predictions (top row), as well as the corresponding SLD profiles (bottom row). By using a FNO as the embedding network, our approach is successfully extended to curves with variable discretizations. A disadvantage in this scenario is that the network takes longer to converge.

\clearpage

\section{Conclusions}

In this study, we address the phase problem as the primary obstacle in machine learning-based approaches for extracting information from X-ray reflectivity (XRR) and neutron reflectivity (NR) data. The lack of phase information induces non-uniqueness in the space of possible solutions, the inverse problem becoming increasingly underdetermined the larger the considered parameter space. This prevents the successful training of neural networks for complex multilayer structures, with many free parameters. Therefore, previous solutions were limited to relatively simple layer structures with few parameters. To tackle this issue, we propose a procedure that enables training a neural network on a continuous range of smaller subspaces of a large parameter space. Our method allows users to incorporate prior experimental knowledge by specifying upper and lower bounds for each parameter during inference. This approach overcomes limitations in existing methods, as it allows training networks with a larger number of parameters and expanded parameter ranges, while still enabling proper convergence. Additionally, we introduce the use of a neural operator to process reflectivity curves with varying discretizations. We validate the effectiveness of our approach by training networks using different physical models: two-layer and five-layer box model parameterizations, as well as a specialized parameterization for repeating identical monolayers. In contrast to previous work, our approach scales favourably when increasing the complexity of the inverse problem, giving good predictions even for the challenging 5-layer multilayer model. We note that the proposed approach can be adapted to tackle other inverse problems in science affected by the non-uniqueness issue.

\section{Author contributions}

    V.M. trained neural networks, performed all the tests, and implemented the FNO technique. V.S. introduced the prior-aware ML approach, designed the neural network, and implemented the training procedure. S.K. and V.S. designed and implemented multilayer parameterization. A.H. performed the experiments. A.Gr., L.P., A.G., A.H., S.K., and F.S. provided expertise in surface scattering. All the authors contributed to writing the paper, analysis, and interpretation of the results. A.H. and F.S. supervised the research process.

\section{Acknowledgements}
This research is part of a project (VIPR, 05D23VT1) funded by the German Federal Ministry for Science and Education (BMBF). This work is partly supported by the consortium DAPHNE4NFDI in the context of the work of the NFDI e.V., funded by the German Research Foundation (DFG). We acknowledge DESY (Hamburg, Germany), a member of the Helmholtz Association, for the provision of experimental facilities.

\bibliography{references_reflectivity}

\end{document}